\documentclass[11pt,a4paper]{article}
\usepackage[margin=24mm]{geometry}
\usepackage[T1]{fontenc}
\usepackage{lmodern}
\usepackage{amsmath,amssymb,booktabs,graphicx,array}
\usepackage[font=small,labelfont=bf]{caption}
\usepackage{microtype}
\usepackage{placeins}
\usepackage[numbers,sort&compress]{natbib}
\usepackage{xurl}
\usepackage{xcolor}
\usepackage{hyperref}
\hypersetup{colorlinks=true,linkcolor=blue!45!black,citecolor=blue!45!black,urlcolor=blue!45!black,pdfauthor={Zhuo Wang, Ce Yu, Bin Yang*}}
\graphicspath{{figures/}}
\newcommand{\XR}{Xu--Randall}
\newcommand{\clip}{\operatorname{clip}}
\newcommand{\logit}{\operatorname{logit}}
\newcommand{\MSE}{\operatorname{MSE}}

\title{Physics-anchored residual quantum neural networks for cloud-cover parameterization}
\author{Zhuo Wang, Ce Yu, Bin Yang*\\[0.3em]\small Tianjin University}
\date{}

\begin{document}
\maketitle

\begin{abstract}
Compact quantum neural networks have been explored for cloud-cover parameterization, but their predictive value depends on how physical knowledge is incorporated and which classical models provide the reference. We investigate an offline residual formulation that retains a fitted \XR\ cloud-cover estimate and learns a correction to its logit in transformed target space. Using released ICON/DYAMOND-derived data, we compare ten model variants in six- and eight-input settings, with three matched replicates and 10,000 training samples per replicate. All quantum circuits use local Z readout and noiseless exact expectation values. Under a common 30-epoch protocol, residual prediction reduced physical-space test mean squared error relative to direct QNN prediction by 54.3\% and 38.8\% for six-input XYZ and ZZXY circuits, and by 49.2\% and 38.9\% for their eight-input versions. Residual QNNs also outperformed the selected compact residual multilayer perceptron in every paired replicate. A larger, longer-trained Grundner-Q1 adaptation nevertheless achieved lower error: residual Q1 reached mean MSEs of 0.00778 and 0.00791, compared with 0.01124 and 0.01082 for residual XYZ. The results support residual prediction for compact simulated QNNs within the tested budget and establish a clear boundary set by stronger classical references. They do not establish computational quantum advantage or online climate-model skill.
\end{abstract}

\section{Introduction}

\subsection{Cloud-cover parameterization and existing learning approaches}

Cloud-cover parameterizations estimate the cloudy fraction of a model grid cell from resolved atmospheric variables. This diagnostic connects unresolved cloud structure to the radiation and microphysics represented in a climate model. Coarse spatial resolution makes the relationship difficult: similar mean atmospheric states can contain different sub-grid distributions of water and cloud. Storm-resolving simulations provide training examples for learning this relationship, although accurate offline predictions do not by themselves establish skill after coupling to atmospheric dynamics \cite{Grundner2022,Stevens2019,Zaengl2015}.

Semiempirical schemes such as \XR\ express cloudiness through a compact physical relationship \cite{XuRandall1996}. Classical neural networks offer more flexible mappings, with Grundner et al. demonstrating learned cloud-cover parameterizations from coarse-grained ICON simulations \cite{Grundner2022}. Related work on unresolved atmospheric processes and scientific machine learning shows how data-driven models can be combined with existing process knowledge \cite{Rasp2018,Reichstein2019,Willard2022}. One consequential choice is whether a network learns the complete diagnostic or corrects a useful approximation. The latter retains information from the approximation and its calibration, changing what the trainable component must represent.

Quantum neural networks (QNNs) provide another compact parameterized function class. Classical inputs enter a circuit as rotation angles, and measured expectation values supply regression features. Repeated encoding, or data re-uploading, changes the accessible input-dependent frequency structure \cite{PerezSalinas2020,Schuld2021}. Pastori et al. applied such circuits to cloud-cover prediction and compared them with classical networks of similar parameter counts \cite{Pastori2026}. Their data, transformations, circuits, and fitted physical baseline define our starting point. Whether a circuit is useful for prediction remains an empirical question requiring appropriate classical references. Precise simulations on classical computers do not prove quantum computing superiority \cite{Cerezo2021,Bowles2024}.

\subsection{Research questions and contributions}

We ask whether a compact QNN benefits from correcting a physical cloud-cover estimate rather than predicting cloud cover directly. A second question is whether any advantage persists when classical models receive the same physical anchor, and when a stronger published classical architecture is included. Finally, we examine whether the findings persist in the paper-defined eight-input configuration on the same sample splits.

We address these questions with a bounded correction in transformed-logit space, direct and residual XYZ and ZZXY circuits, compact classical controls, and direct and residual Grundner-Q1 adaptations. The contribution is an evaluated prediction design and a matched comparison with explicit limits. Direct and residual circuits have identical trainable parameter counts; all residual controls receive the same inherited baseline. Six- and eight-input comparisons share sample manifests. This design preserves the distinction between near-sized comparisons under a shared training protocol and stronger classical references trained under a different protocol.

\section{Methods}

\subsection{Xu--Randall anchoring and residual prediction}

Let $x$ denote a cell's atmospheric inputs and $c\in[0,1]$ its physical cloud fraction. We use the fitted \XR-derived diagnostic supplied by the reference implementation \cite{XuRandall1996,Pastori2026}:
\begin{equation}
 c_{\mathrm{XR}}(x)=\min\!\left\{1,\left[1-e^{-\alpha(q_l+q_i)}\right]\mathrm{RH}^{\beta}\right\},
 \quad \alpha=4.034\times10^4,\quad\beta=0.9942.
 \label{eq:xr}
\end{equation}
Here $q_l$ and $q_i$ are liquid-water and ice specific contents, and relative humidity is diagnosed from specific humidity, temperature, and pressure. The two constants remain fixed. Their earlier calibration supplies information beyond the selected training subset; we report them separately from newly trainable parameters. Equation~\eqref{eq:xr} records the implementation used here, rather than a new fit of the complete original 1996 scheme.

\begin{figure}[tbp]
\centering
% Preserve the audited diagram's minimum effective type size above 5 pt.
\makebox[\textwidth][c]{\includegraphics[width=164mm]{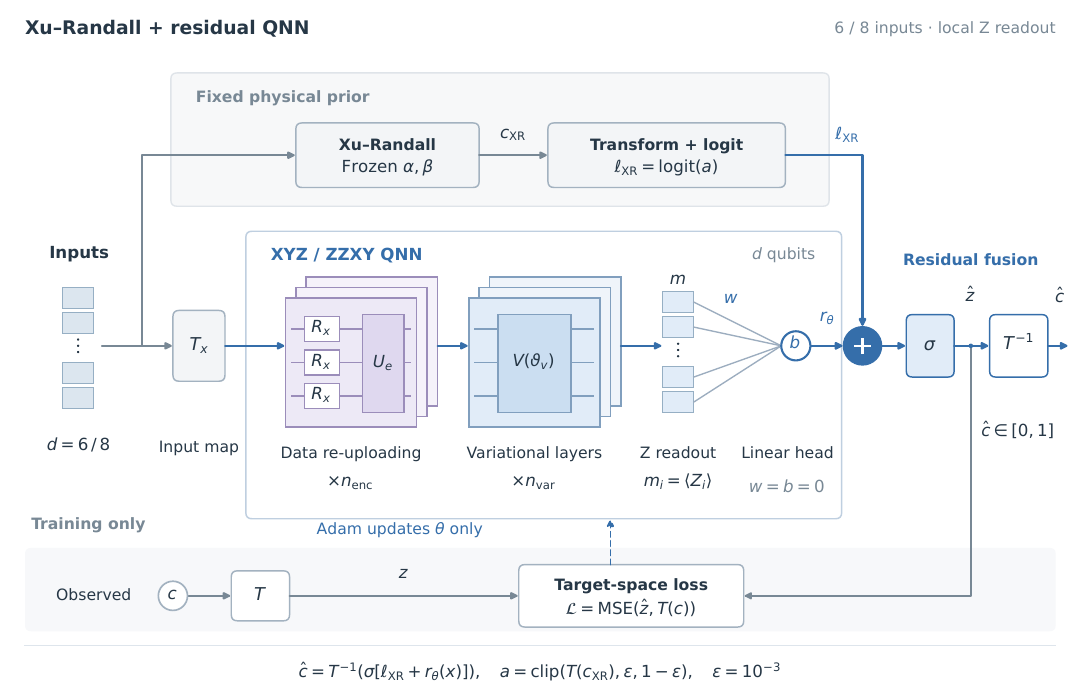}}
\caption{Physics-anchored residual QNN. The fixed branch supplies $\logit a(x)$. The quantum branch supplies local expectations $m_i=\langle Z_i\rangle$ to one affine head. Its zero initialization recovers the clipped anchor. The branches combine before the sigmoid, and the inverse target transform returns physical cloud cover. Training loss is evaluated before that inverse transform; the dashed arrow updates only trainable parameters. Overlapping blocks and representative wires abbreviate the six- or eight-qubit circuit, not its actual layer count. The figure's $T$ and $T_x$ denote the target and input transforms.}
\label{fig:method}
\end{figure}

With the inherited target transform $z=\mathcal{T}(c)$, define
\begin{equation}
 a(x)=\clip\!\left(\mathcal{T}(c_{\mathrm{XR}}(x)),\epsilon,1-\epsilon\right),
 \qquad \epsilon=10^{-3}.
 \label{eq:anchor}
\end{equation}
A trainable correction $r_\theta(x)$ produces
\begin{equation}
 \hat z(x)=\sigma\!\left[\logit a(x)+r_\theta(x)\right],
 \qquad \hat c(x)=\mathcal{T}^{-1}(\hat z(x)),
 \label{eq:residual}
\end{equation}
where $\sigma(s)=(1+e^{-s})^{-1}$. The correction is additive in the logit of the transformed target, rather than in physical cloud fraction. Figure~\ref{fig:method} shows the fixed physical branch and trainable quantum branch.

Every residual output head starts at zero, recovering $a(x)$ up to numerical precision while retaining random circuit or hidden-layer initialization. Clipping keeps the anchor logit finite at the endpoints. We minimize $N^{-1}\sum_i[\hat z_i-\mathcal{T}(c_i)]^2$ and evaluate after inverse transformation. Observed targets are never converted to logits, so zero and one remain valid finite training targets. The sigmoid bounds predictions without constraining the correction to be small; the transformed loss weights physical errors differently from physical-space MSE.

\subsection{Quantum circuits and local Z readout}

We reuse the XYZ and ZZXY implementations with one qubit per input and noiseless exact expectations. Each input is repeatedly encoded through an $R_X$ rotation. XYZ encoding blocks contain nearest-neighbor Ising ZZ, XX, and YY operations; its subsequent variational blocks additionally apply local $R_X$ rotations. ZZXY encoding blocks combine Ising ZZ interactions with local $R_Y$ rotations, followed by variational blocks containing local $R_X$, Ising ZZ, and local $R_Y$ operations. Encoding and variational layer counts $(n_{\mathrm{enc}},n_{\mathrm{var}})$ are $(4,2)$ and $(2,5)$ for six-input XYZ and ZZXY, and $(5,3)$ and $(2,7)$ for eight inputs.

Both prediction modes use the same local observables and affine readout:
\begin{equation}
 m_i(x;\vartheta)=\langle 0|U_\vartheta^\dagger(x)Z_iU_\vartheta(x)|0\rangle,
 \qquad h_\theta(x)=b+\sum_{i=1}^{d}w_i m_i(x;\vartheta).
\end{equation}
The direct model uses $h_\theta$ to estimate $z$; the residual model uses it as $r_\theta$ in Eq.~\eqref{eq:residual}. Local Z readout concerns measured observables and does not remove the circuit's two-qubit operations. Circuit angles and affine coefficients give 109/114 trainable parameters for six-input XYZ/ZZXY and 201/200 for eight inputs, unchanged between direct and residual versions. These counts describe the regression model, not the memory or execution cost of its state-vector simulator.

\subsection{Classical residual controls and the Grundner Q1 adaptation}

The compact direct multilayer perceptron (MLP) follows the reference structures $6\!\to\!8\!\to\!3\!\to\!7\!\to\!1$ and $8\!\to\!12\!\to\!6\!\to\!2\!\to\!1$, with tanh hidden layers and linear outputs. Its residual counterpart preserves the hidden layers and supplies Eq.~\eqref{eq:residual} through a zero-initialized final layer. A linear residual control uses $r_\phi(x)=w^\top x+b$, with all coefficients initially zero. These models test whether simple classical corrections share the benefit and provide a compact nonlinear reference.

The stronger reference adapts Grundner's cell-based Q1 architecture \cite{Grundner2022}: $d\!\to\!64\!\to\!64\!\to\!64\!\to\!1$, with tanh, LeakyReLU of slope 0.2, and tanh hidden activations. Batch normalization follows the second activation. Every Dense kernel has L1 and L2 penalties of 0.004749 and 0.008732. The network has 128 non-trainable normalization states in addition to Table~\ref{tab:models}'s parameters. Direct Q1 uses seeded Glorot kernels and zero biases; residual Q1 also zeroes the final kernel. This adapts the published architecture to the present features, targets, and sample size.

Q1 uses a numerical target scale of 100 to retain the scale of its inherited regularization. Its direct output $g_\theta$ minimizes $\MSE(g_\theta,100z)+L_{\mathrm{reg}}$ and predicts $g_\theta/100$. Its residual output supplies the dimensionless logit correction and minimizes $\MSE(100\hat z,100z)+L_{\mathrm{reg}}$. These losses use the transformed target, not the original Q1 paper's physical percentage target.

\begin{table}[tbp]
\centering\small
\caption{Trainable parameters and training protocols. Every residual also uses two fixed \XR\ constants; Q1 adds 128 non-trainable normalization states. Compact models are near-sized across families and exactly size-matched within each direct/residual pair.}
\label{tab:models}
\begin{tabular}{lrrl}
\toprule
Model family & Six inputs & Eight inputs & Training protocol \\
\midrule
\XR & 0 & 0 & Fixed calibration \\
Linear residual & 7 & 9 & Adam, 30 epochs \\
Compact MLP, direct or residual & 119 & 203 & Adam, 30 epochs \\
XYZ QNN, direct or residual & 109 & 201 & Adam, 30 epochs \\
ZZXY QNN, direct or residual & 114 & 200 & Adam, 30 epochs \\
Q1, direct or residual & 8,961 & 9,089 & Adam, at most 300 epochs \\
\bottomrule
\end{tabular}
\end{table}

\section{Experimental design}

\subsection{Data, input configurations, and matched replicates}

We use the released cell-based ICON/DYAMOND-derived arrays accompanying Pastori et al. \cite{Pastori2026}, without rerunning simulations or coarse-graining. Our loader retains rows with $q_l+q_i>10^{-8}\,\mathrm{kg\,kg^{-1}}$; evaluation is conditional on this condensate-filtered distribution. The six inputs are specific humidity $q_v$, liquid water $q_l$, ice $q_i$, temperature $T_a$, pressure $p$, and horizontal wind speed $h_w=\sqrt{u^2+v^2}$. Eight-input order is $(q_v,q_l,q_i,T_a,p,z_g,h_w,f)$, adding geometric height and the Coriolis parameter. Compact models use the inherited fixed input transforms; Q1 uses raw selected variables standardized using training-set means and population standard deviations only. All models share $\mathcal{T}$ (Appendix~\ref{app:implementation}).

Each of three replicates contains 10,000 training, 2,000 validation, and 20,000 test cells. Training and validation rows are disjoint selections from the released training shards; test rows come from testing shards. Sampling approximately balances 64 shards in each pool. Root seed 12345 generates separate data, initialization, and shuffle seeds. Models share rows and batch permutations within a replicate; six- and eight-input settings use byte-identical split manifests, verified by SHA-256. These matched replicates vary both selected rows and initialization and are not temporal cross-validation folds. Test sets across replicates can overlap. The ten models in each setting yield 60 evaluation records, including six fixed-baseline evaluations. The \XR\ diagnostic continues to use its original five inputs in both settings.

\subsection{Training budgets, initialization, and checkpoint selection}

Compact trainable models use Adam \cite{Kingma2014}, learning rate $10^{-3}$, epsilon $10^{-7}$, batch size 100, and 30 epochs. QNN angles are initialized uniformly on $[-\pi/2,\pi/2]$ and shared across a circuit's direct/residual pair within each replicate. Direct heads are randomly initialized; residual heads start at zero. Compact MLPs retain the reference random initialization except for their residual final layer.

Q1 uses Adam with learning rate $4.33\times10^{-4}$, epsilon 0.1, batch size 100, and at most 300 epochs. Its early-stopping counter resets after improvement exceeding $10^{-6}$ in unregularized validation MSE, with patience 30. Independently of this tolerance, each trainer saves the earliest epoch attaining the exact minimum validation MSE. Only that checkpoint is tested. Direct transformed predictions are clipped to $[0,1]$ before inversion. The shared compact-model protocol and longer Q1 protocol answer different comparison questions; neither stopping rule certifies optimization convergence.

\subsection{Evaluation metrics and paired statistics}

Physical-space test MSE is the primary metric. We also report $R^2=1-\sum_i(\hat c_i-c_i)^2/\sum_i(c_i-\bar c)^2$, histogram Hellinger distance with 100 equal bins on $[0,1]$, and empirical one-dimensional Wasserstein distance. MSE and $R^2$ assess sample-wise errors and are reported in Table~\ref{tab:results}; the distributional distances compare marginal physical cloud-fraction distributions and are reported separately in Table~\ref{tab:distributions}. Lower values are better for both distances, whereas higher $R^2$ is better. On the same test set, MSE and $R^2$ give the same ranking.

We report actual replicate values, means, sample standard deviations, and win counts. Percentage improvements are calculated within replicate $k$ as $100(\MSE_{\mathrm{ref},k}-\MSE_{\mathrm{cand},k})/\MSE_{\mathrm{ref},k}$ before averaging, rather than taking a ratio of mean MSEs. Three repeats support descriptive comparisons, not significance claims or uncertainty estimates over independent climate realizations. The manuscript data audit recomputes metrics from saved predictions and checks aligned targets, manifests, residual reconstruction, and selected validation minima.

\section{Results}

\begin{table}[tbp]
\centering\scriptsize
\caption{Physical-space performance, mean $\pm$ sample SD over three matched replicates of 20,000 test cells each. All QNNs use local Z readout. Compact models share 30 epochs; Q1 uses its separate longer protocol.}
\label{tab:results}
\begin{tabular}{lrrrr}
\toprule
Model & \multicolumn{2}{c}{Six inputs} & \multicolumn{2}{c}{Eight inputs} \\
\cmidrule(lr){2-3}\cmidrule(lr){4-5}
 & MSE $\downarrow$ & $R^2$ $\uparrow$ & MSE $\downarrow$ & $R^2$ $\uparrow$ \\
\midrule
Xu--Randall & $0.04804\pm0.00067$ & $0.504\pm0.008$ & $0.04804\pm0.00067$ & $0.504\pm0.008$ \\
Compact MLP, direct & $0.04277\pm0.00599$ & $0.558\pm0.062$ & $0.02753\pm0.00813$ & $0.715\pm0.085$ \\
XYZ QNN, direct & $0.02547\pm0.00530$ & $0.737\pm0.054$ & $0.02165\pm0.00374$ & $0.776\pm0.038$ \\
ZZXY QNN, direct & $0.02002\pm0.00329$ & $0.793\pm0.034$ & $0.01815\pm0.00168$ & $0.812\pm0.017$ \\
XR + linear & $0.01663\pm0.00008$ & $0.828\pm0.000$ & $0.01667\pm0.00014$ & $0.828\pm0.001$ \\
XR + compact MLP & $0.01322\pm0.00039$ & $0.863\pm0.004$ & $0.01310\pm0.00017$ & $0.865\pm0.002$ \\
XR + XYZ QNN & $0.01124\pm0.00017$ & $0.884\pm0.002$ & $0.01082\pm0.00022$ & $0.888\pm0.002$ \\
XR + ZZXY QNN & $0.01203\pm0.00012$ & $0.876\pm0.001$ & $0.01101\pm0.00044$ & $0.886\pm0.005$ \\
Q1, direct & $0.01003\pm0.00055$ & $0.896\pm0.005$ & $0.00941\pm0.00041$ & $0.903\pm0.004$ \\
XR + Q1 & $0.00778\pm0.00006$ & $0.920\pm0.001$ & $0.00791\pm0.00008$ & $0.918\pm0.001$ \\
\bottomrule
\end{tabular}

\end{table}

\begin{figure}[tbp]
\centering
\includegraphics[width=\textwidth]{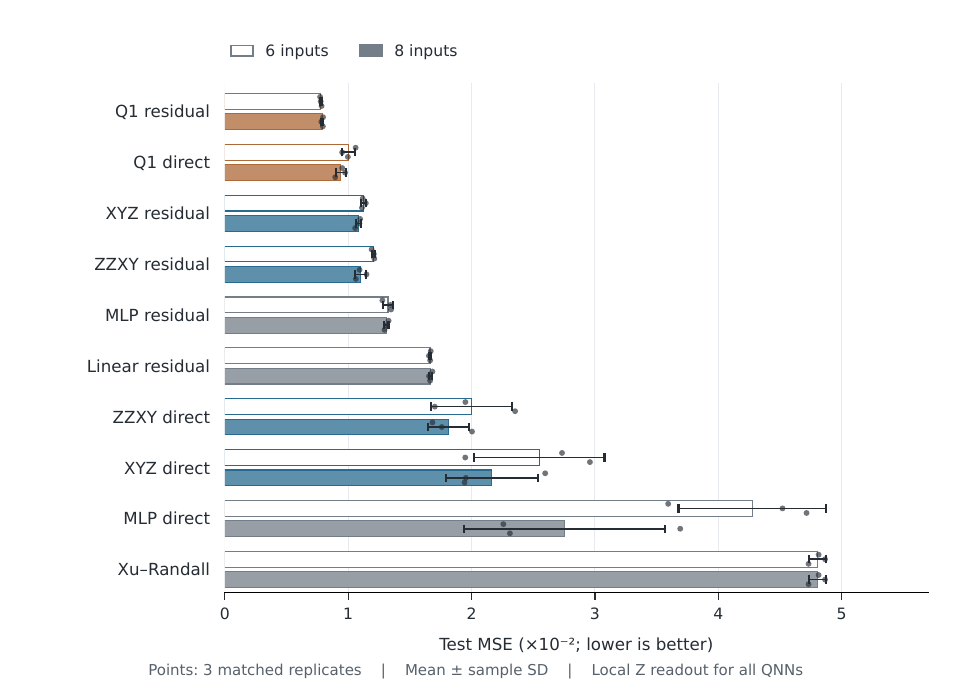}
\caption{Test-set performance across input configurations. Each row groups six-input (open) and eight-input (filled) bars, whose lengths show mean physical-space MSE. Small points represent three matched replicates; whiskers show $\pm$ one sample SD ($n=3$, 20,000 test cells per replicate). QNNs are blue, Q1 adaptations ochre, and remaining controls gray. Every residual model is anchored to \XR; all QNNs use local Z readout. Q1 uses a larger architecture and a separate, longer training protocol. Axis values are expressed in units of $10^{-2}$; lower values indicate lower error. Source data: \texttt{runs.csv} and \texttt{summary.csv}.}
\label{fig:performance}
\end{figure}

\subsection{Residual prediction improves both compact QNN architectures}

\begin{figure}[tbp]
\centering
\includegraphics[width=\textwidth]{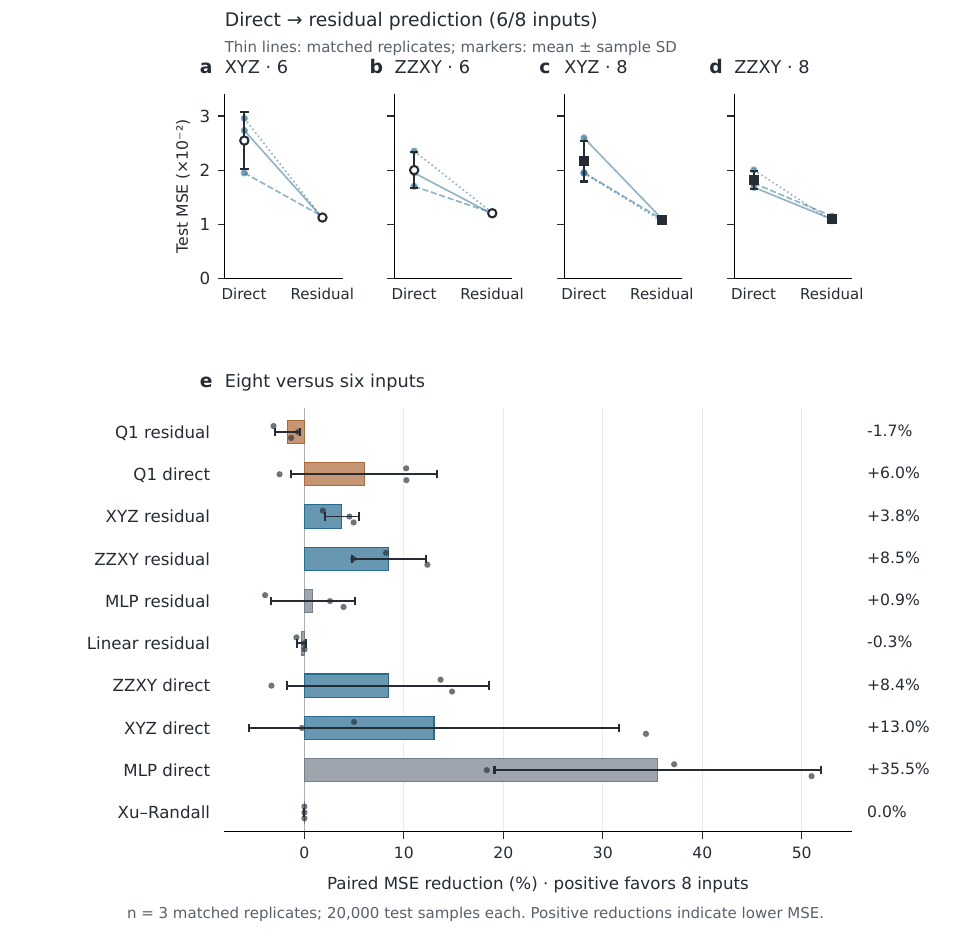}
% \caption{Paired effects of residual prediction and input configuration. \textbf{a--d}, direct-to-residual QNN comparisons for six-input XYZ, six-input ZZXY, eight-input XYZ, and eight-input ZZXY, respectively. Each thin line connects one matched replicate; small endpoints show its physical-space MSE. Large open circles (six inputs) and filled squares (eight inputs) show means, with whiskers indicating $\pm$ one sample SD. \textbf{e}, percentage MSE reductions for eight inputs relative to six, computed within replicate before averaging. Bars and right-hand labels show mean reductions; small points show individual pairs, and whiskers indicate $\pm$ one sample SD. Positive values favor eight inputs; colors identify QNNs (blue), Q1 (ochre), and other controls (gray). All panels use three matched replicates with 20,000 test cells each. QNN inputs, qubit count, and circuit depth change together in panel e. Source data: \texttt{runs.csv}, \texttt{paired\_contrasts.csv}, and \texttt{feature\_pairs.csv}.}
\caption{Paired effects of residual prediction and input configuration. \textbf{a--d}, direct-to-residual QNN comparisons for six-input XYZ, six-input ZZXY, eight-input XYZ, and eight-input ZZXY, respectively. Each thin line connects one matched replicate; small endpoints show its physical-space MSE. Large open circles (six inputs) and filled squares (eight inputs) show means, with whiskers indicating $\pm$ one sample SD. \textbf{e}, percentage MSE reductions for eight inputs relative to six, computed within replicate before averaging. Bars and right-hand labels show mean reductions; small points show individual pairs, and whiskers indicate $\pm$ one sample SD. Positive values favor eight inputs; colors identify QNNs (blue), Q1 (ochre), and other controls (gray). All panels use three matched replicates with 20,000 test cells each. QNN inputs, qubit count, and circuit depth change together in panel e.}
\label{fig:paired}
\end{figure}

Residual prediction reduced test MSE in all twelve matched comparisons across the two circuits and input settings (Fig.~\ref{fig:paired}a--d; Table~\ref{tab:results}). With six inputs, XYZ changed from mean MSE 0.02547 to 0.01124 and ZZXY from 0.02002 to 0.01203. The mean paired reductions were $54.3\pm11.3\%$ and $38.8\pm9.7\%$. With eight inputs, the corresponding reductions were $49.2\pm7.5\%$ and $38.9\pm7.0\%$, without adding trainable parameters within a circuit family.

Lower sample-wise MSE did not imply uniformly better distributional agreement (Table~\ref{tab:distributions}; lower is better for both distances). For eight-input XYZ, residual prediction increased mean Hellinger distance from 0.0713 to 0.0802, indicating worse agreement by this measure, but decreased mean Wasserstein distance from 0.0220 to 0.0197, indicating better agreement. The fixed \XR\ baseline also had lower mean Wasserstein distance than several learned models despite much larger MSE (Table~\ref{tab:results}). These distances compare the overall distributions without retaining the correspondence between individual predictions and targets. The residual benefit therefore concerns sample-wise prediction accuracy; the distributional diagnostics did not improve uniformly.

\begin{table}[tbp]
\centering\footnotesize
\setlength{\tabcolsep}{3.5pt}
% \caption{Distributional distances in physical cloud-fraction space. Values are means $\pm$ sample SD over three matched replicates, with 20,000 test cells per replicate. Both Hellinger and Wasserstein distances are lower-is-better ($\downarrow$), unlike $R^2$. Hellinger uses 100 equal bins on $[0,1]$; Wasserstein is the empirical one-dimensional distance. Models and training protocols match Table~\ref{tab:results}; all QNNs use local Z readout. Source data: \texttt{runs.csv} and \texttt{summary.csv}.}
\caption{Distributional distances in physical cloud-fraction space. Values are means $\pm$ sample SD over three matched replicates, with 20,000 test cells per replicate. Both Hellinger and Wasserstein distances are lower-is-better ($\downarrow$), unlike $R^2$. Hellinger uses 100 equal bins on $[0,1]$; Wasserstein is the empirical one-dimensional distance. Models and training protocols match Table~\ref{tab:results}. All QNNs use local Z readout.}
\label{tab:distributions}
\begin{tabular}{lrrrr}
\toprule
Model & \multicolumn{2}{c}{Six inputs} & \multicolumn{2}{c}{Eight inputs} \\
\cmidrule(lr){2-3}\cmidrule(lr){4-5}
 & Hellinger $\downarrow$ & Wasserstein $\downarrow$ & Hellinger $\downarrow$ & Wasserstein $\downarrow$ \\
\midrule
Xu--Randall & $0.0972\pm0.0011$ & $0.0189\pm0.0012$ & $0.0972\pm0.0011$ & $0.0189\pm0.0012$ \\
Compact MLP, direct & $0.2320\pm0.0655$ & $0.0716\pm0.0141$ & $0.2130\pm0.0902$ & $0.0339\pm0.0022$ \\
XYZ QNN, direct & $0.0800\pm0.0084$ & $0.0247\pm0.0018$ & $0.0713\pm0.0074$ & $0.0220\pm0.0019$ \\
ZZXY QNN, direct & $0.0908\pm0.0065$ & $0.0279\pm0.0015$ & $0.0774\pm0.0096$ & $0.0249\pm0.0014$ \\
XR + linear & $0.0865\pm0.0023$ & $0.0246\pm0.0008$ & $0.0845\pm0.0026$ & $0.0242\pm0.0009$ \\
XR + compact MLP & $0.0993\pm0.0045$ & $0.0289\pm0.0013$ & $0.0857\pm0.0103$ & $0.0233\pm0.0035$ \\
XR + XYZ QNN & $0.0831\pm0.0061$ & $0.0214\pm0.0032$ & $0.0802\pm0.0059$ & $0.0197\pm0.0012$ \\
XR + ZZXY QNN & $0.0860\pm0.0019$ & $0.0237\pm0.0007$ & $0.0827\pm0.0060$ & $0.0225\pm0.0023$ \\
Q1, direct & $0.1624\pm0.0115$ & $0.0140\pm0.0025$ & $0.1199\pm0.0146$ & $0.0130\pm0.0036$ \\
XR + Q1 & $0.0819\pm0.0045$ & $0.0127\pm0.0025$ & $0.0785\pm0.0006$ & $0.0112\pm0.0014$ \\
\bottomrule
\end{tabular}

\end{table}

\subsection{Classical residual controls also benefit, while QNNs outperform the selected compact MLP}

The compact residual MLP also improved in every replicate (Fig.~\ref{fig:performance}). Its mean MSE fell from 0.04277 to 0.01322 with six inputs and from 0.02753 to 0.01310 with eight. Even the seven-parameter linear correction attained MSE 0.01663; its nine-parameter version attained 0.01667. These results establish that useful corrections to the inherited baseline can be learned with classical models.

Within the shared 30-epoch protocol, residual QNNs nevertheless outperformed the selected residual MLP in every pair. XYZ reduced its MSE by $14.9\pm2.5\%$ with six inputs and $17.4\pm0.9\%$ with eight; ZZXY reductions were $9.0\pm1.9\%$ and $16.0\pm3.2\%$. Parameter counts were comparable but not identical across families (Table~\ref{tab:models}). The result favors these circuits over this compact classical control under the specified protocol, without establishing superiority over the class of similarly sized classical models.

\subsection{The stronger Grundner Q1 reference remains more accurate}

Both Q1 adaptations achieved lower MSE than residual XYZ in every matched replicate (Fig.~\ref{fig:performance}). Direct Q1 attained means of $0.01003\pm0.00055$ and $0.00941\pm0.00041$ for six and eight inputs. Residual Q1 improved further to $0.00778\pm0.00006$ and $0.00791\pm0.00008$, with mean $R^2$ of 0.920 and 0.918. Its paired MSE reductions relative to residual XYZ were $30.8\pm1.3\%$ and $26.8\pm1.9\%$.

Residual Q1 also beat direct Q1 in every replicate. Its advantage reinforces the usefulness of residual prediction while bounding the quantum-model comparison. Q1 differs in capacity, preprocessing, regularization, optimizer settings, and training duration. Selected direct Q1 epochs ranged from 110--134 for six inputs and 139--300 for eight; one run selected the training cap. Residual Q1 selected epochs 127--219 and 167--210. These are empirical accuracy references under their protocols, not a classical performance ceiling or an isolated test of network width.

\subsection{Benefits of the eight-input configuration depend on the model}

On matched splits, eight-input residual XYZ reduced mean MSE from 0.01124 to 0.01082, a $3.8\pm1.7\%$ mean paired improvement (Fig.~\ref{fig:paired}e). Residual ZZXY improved from 0.01203 to 0.01101, corresponding to a mean paired reduction of $8.5\pm3.7\%$. Both circuits improved in all three replicates, whereas each direct QNN improved in only two despite lower mean MSE. Benefits were not uniform across families: linear residual MSE increased slightly, and residual Q1 was approximately 1.7\% worse. Six inputs were preferred in all three residual Q1 pairs.

These differences describe complete configurations. Eight-input QNNs change the variables, qubit count, and paper-specified depth together; the compact MLP shape also changes. Consequently, improvements cannot be assigned to geometric height or the latitude-related input alone. The unchanged \XR\ predictions confirm the aligned baseline, while the model-dependent outcomes delimit the benefit of the extended configuration.

\FloatBarrier
\section{Discussion}

\subsection{Interpreting physical residual prediction}

The results support retaining a calibrated physical predictor within a compact QNN under the tested training budget. The benefit across classical residual controls places this finding within a broader strategy of correcting existing approximations. It extends the direct cloud-cover setting of Pastori et al. \cite{Pastori2026}; its usefulness does not depend on claiming that the correction strategy is exclusive to quantum circuits. Although residual QNNs outperform the selected near-sized MLP, stronger Q1 references change the ranking.

This construction uses physical knowledge through a fixed predictor, rather than governing-equation residuals in a physics-informed loss \cite{Karniadakis2021}. Bounded output also differs from enforcing analytic conservation constraints \cite{Beucler2021}. A learned correction can depart substantially from its anchor, and inherited calibration adds information beyond the 10,000 selected training examples. The experiment therefore establishes neither full physical consistency nor sample-efficiency scaling from an equal total-information budget.

The observed gain is attributable to the complete residual design. Relative to direct prediction, it changes the anchor, sigmoid output, and zero-head initialization simultaneously. The derivative $\partial\hat z/\partial r=\hat z(1-\hat z)$ can attenuate gradients near the boundaries; zero head weights also make initial circuit-angle gradients zero, although head gradients can be nonzero. These properties suggest possible optimization effects but were not isolated experimentally. Holding the output map and initialization fixed while removing or replacing the physical anchor is the next discriminating test of its specific contribution.

\subsection{Scope and next discriminating tests}

The comparison covers one training size, three matched repeats, and particular architectures. Many compact-model validation histories continue improving near epoch 30. Q1's longer protocol supplies an accuracy reference, but cannot isolate a capacity effect. A comparison with matched optimization budgets and stronger compact classical tuning would test how far the observed circuit-specific benefit persists \cite{Bowles2024}.

The condensate filter and shard-balanced sampling further limit the evaluated distribution. Random cells from the released pools do not constitute new temporal or geographical blocking, and shared weather conditions can induce dependence across samples and repeats. Evaluating matched protocols on blocked periods or regions is needed to assess portability beyond the present split design. Input-number comparisons additionally remain confounded with depth and architecture changes.

Finally, all quantum predictions are exact state-vector expectations computed on classical accelerators. Parameter compactness does not measure simulation, encoding, measurement, or gradient costs. Finite shots, hardware noise, and hardware speed were not tested, and no online atmospheric coupling was performed. The supported outcome is an offline predictive improvement under specified conditions; computational advantage and climate-model feedback require separate evidence.

\section{Conclusion}

A bounded logit correction around the fitted \XR\ anchor improved local-Z-readout XYZ and ZZXY QNNs in every matched replicate of the six- and eight-input experiments. These compact residual QNNs also surpassed the selected residual MLP under the shared 30-epoch protocol, while the larger Grundner-Q1 adaptations remained more accurate. The contribution is a reproducible prediction design and comparison that identifies both useful gains and the limits imposed by stronger classical references.

% \section*{Data and code availability}
% Data and upstream circuits are described by Pastori et al. \cite{Pastori2026} and their \href{https://github.com/EyringMLClimateGroup/pastori26MLEarth_QNN-cloudcover}{reference repository}, which identifies dataset DOI \href{https://doi.org/10.5281/zenodo.21455691}{10.5281/zenodo.21455691}. The manuscript source bundle includes audited metrics, paired contrasts, histories, and provenance hashes. \authorinput{confirm dataset accessibility and provide the permanent release URL for the new residual implementation, predictions, and weights.}

% \section*{Declarations}
% \authorinput{contributions, funding, competing interests, and acknowledgments.} This working draft and its preparation code used AI assistance with the supplied experimental artifacts. \authorinput{verify the scientific interpretation and references, and approve the final assistance disclosure.}

\appendix
\section{Exact implementation and supplementary results}\label{app:implementation}

\subsection{Transforms, physical baseline, and initialization}
The inherited target transform and inverse are
\begin{equation}
 \begin{aligned}
 \mathcal{T}(c)&=\pi^{-1}\arcsin\!\left[2\left(\frac{e^{Bc^A}-1}{e^B-1}\right)^C-1\right]+\tfrac12,\\
 \mathcal{T}^{-1}(z)&=\left[\frac{\log\{1+(e^B-1)u(z)^{1/C}\}}{B}\right]^{1/A},
 \quad u(z)=\tfrac12[1+\sin\{\pi(z-\tfrac12)\}],
 \end{aligned}
\end{equation}
with $A=1.29407913$, $B=-3.20011015$, and $C=0.70308237$. For humidity, liquid water, ice, and wind, define $L(q)=\log[1+(e-1)(q/q_{\max})^\gamma]$ and input angle $\pi[L(q)-L(q_{\min})]/[1-L(q_{\min})]$. The triples $(\gamma,q_{\min},q_{\max})$ are $(0.25,10^{-7},0.025)$, $(0.25,0,0.00145)$, $(0.25,0,0.00055)$, and $(0.5,0.0015,115)$, respectively; moisture units are $\mathrm{kg\,kg^{-1}}$ and wind units $\mathrm{m\,s^{-1}}$. Linear maps send temperature, pressure, and height ranges $(180,305)$ K, $(4000,105000)$ Pa, and $(0,20000)$ m to $[0,\pi]$. Coriolis angle is $\arcsin[f/(2\Omega)]+\pi/2$, with $\Omega=7.2921\times10^{-5}\,\mathrm{s^{-1}}$. Input angles outside fitted ranges are not clipped.

For Eq.~\eqref{eq:xr}, $\mathrm{RH}=\max(e/e_s,0)$ and $e=pq_v/(0.622+0.378q_v)$. Saturation pressures are $e_w=611.21\exp[17.269(T_a-273.15)/(T_a-35.86)]$ and $e_i=611.21\exp[21.875(T_a-273.15)/(T_a-7.66)]$ Pa. Between 250.15 and 273.15 K, $e_s=e_i+(e_w-e_i)[(T_a-250.15)/23]^2$; outside, $e_i$ or $e_w$ applies. Fixed transforms and calibration are inherited. Zero residual heads recover Eq.~\eqref{eq:anchor}; clipping changed physical anchor values by less than $2\times10^{-5}$ on the tests.

% \subsection{Secondary metrics and training diagnostics}
% The source tables retain all 60 records, secondary metrics, selected epochs, and timings. Full validation histories and the supplementary learning-curve figure are included with the source bundle. The audit recomputes prediction metrics, reconstructs residual outputs, and checks that each selected epoch is the earliest validation minimum; it records artifact hashes and matching split hashes. Hellinger is $\sqrt{\tfrac12\sum_b(\sqrt{p_b}-\sqrt{q_b})^2}$ for normalized histogram masses; Wasserstein uses empirical samples. The original environments used PennyLane 0.35.1, JAX 0.4.30, Optax 0.2.3, and TensorFlow 2.15.0. The archived commit alone does not identify uncommitted experimental additions; a permanent implementation release remains required.

\begingroup
\small
\setlength{\bibsep}{3pt}
\bibliographystyle{unsrtnat}
\bibliography{references}
\endgroup
\end{document}